\documentclass[aps,prl,twocolumn,showpacs,superscriptaddress,groupedaddress]{revtex4}

\usepackage{bbm}

\usepackage{graphicx}
\usepackage{dcolumn}
\usepackage{amsmath}
\usepackage{bm}
\usepackage{color}
\usepackage{mathrsfs}
\usepackage{epstopdf}
\usepackage{amsmath}
\usepackage{amssymb}
\usepackage{subfigure}
\usepackage{enumitem}
\usepackage{multirow}
\usepackage{exscale}
\usepackage{relsize}

\newcommand{\be}{\begin{equation}}
\newcommand{\ee}{\end{equation}}
\newcommand{\bey}{\begin{eqnarray}}
\newcommand{\eey}{\end{eqnarray}}
\newcommand{\bw}{\begin{widetext}}
\newcommand{\ew}{\end{widetext}}

\newcommand{\ba}{\begin{array}}
\newcommand{\ea}{\end{array}}
\newcommand{\bi}{\begin{itemize}}
\newcommand{\ei}{\end{itemize}}
\newcommand{\bem}{\begin{enumerate}}
\newcommand{\eem}{\end{enumerate}}

\begin{document}

 \title {Group-theoretical approach to the calculation of quantum work distribution
 }

\author{Zhaoyu Fei}
\affiliation{School of Physics, Peking University, Beijing 100871, China}

\author{H. T. Quan} \email[Email: ]{htquan@pku.edu.cn}
\affiliation{School of Physics, Peking University, Beijing 100871, China}
\affiliation{Collaborative Innovation Center of Quantum Matter, Beijing 100871, China}

 \date{\today}

 \begin{abstract}
   %Quantum systems with quadratic Hamiltionians are ubiquitous in physics. Examples include the harmonic oscillator, some lattice models, phonons and Bloch electrons in condensed matter, identical particles in an external potential, some models in quantum field theory and the physical models within the mean-field approximation.
   Usually the calculation of work distributions in an arbitrary nonequilibrium process in a quantum system, especially in a quantum many-body system is extremely cumbersome. For all quantum systems described by quadratic Hamiltonians, we invent a universal method for solving the work distribution of quantum systems in an arbitrary driving process by utilizing the group-representation theory. This method enables us to efficiently calculate work distributions where previous methods fail. In some specific models, such as the time-dependent harmonic oscillator, the dynamical Casimir effect, and the transverse XY model, the exact and analytical solutions of work distributions in an arbitrary nonequilibrium process are obtained. Our work initiates the study of quantum stochastic thermodynamics based on group-representation theory.

 \end{abstract}

 %\pacs{05.45.Mt; 05.45.Pq; 03.67.-a; 64.70.Tg}

 \maketitle

\textit{Introduction}.---In the past two decades, a great breakthrough in the field of nonequilibrium thermodynamics is the discovery of fluctuation theorems~\cite{eq2011} and the establishment of stochastic thermodynamics~\cite{st2010,st2012}. It extends the usual definition of work, heat and entropy production in a thermodynamic process from ensemble-averaged quantities to trajectory-dependent quantities. Based on these extensions, the second law is sharpened and rewritten into equalities (fluctuation theorems) for arbitrary nonequilibrium processes. For an isolated quantum system, a proper definition of the trajectory work is defined through the so-called two-point measurement~\cite{qu2015,pa2018}, which preserves the non-negativity of the probabilities of trajectories and Jarzynski equality at the same time~\cite{aq2000,ja2000}. As is known, the work distribution of a system provides a great deal of information about a nonequilibrium process~\cite{ge2012,th2018,as2015}, which is an analogue to the partition function encoding essential information about an equilibrium state. However, very few analytical results of work distributions of quantum systems have been obtained so far. In literature, the few exceptions are associated with either specific models (e.g., a forced harmonic oscillator~\cite{sta2008,qu2017,sta2019}, a harmonic oscillator with a time-dependent frequency~\cite{no2008,wo2013}, quenched Luttinger liquid at zero temperature~\cite{ge2012}, a diving scalar field~\cite{wor2019} or a transverse Ising model at zero temperature~\cite{wo2013}) or very special protocols (e.g., the sudden quench protocol~\cite{em2012,no2011,fu2019,st2008,ex2015,wo2019,wor2016} or the quantum adiabatic protocol~\cite{wo2012,st2013}). Hence, to develop a universal method to efficiently calculate the work distribution for various quantum systems in an arbitrary nonequilibrium process is one of the most challenging problems in this field.

In this letter, we invent a universal method which is based on the Lie-group theory to calculate the work distribution for quantum systems described by quadratic Hamiltonians~\cite{qu1986,in2010}. Such Hamiltonian is ubiquitous in physics. Examples include the harmonic oscillator, some lattice models, phonons and Bloch electrons in solids, identical particles in an external potential, some models in quantum field theory~\cite{wor2019} and many other physical models within the mean-field approximation, e.g., superconductors and superfluids~\cite{in2010}. The results of work distributions can be solved exactly and efficiently for an arbitrary work protocol. And in many cases, the analytical solution can be obtained explicitly, which not only has pedagogical value, but also brings important insights to the design of optimal protocols for thermal machines~\cite{op2009}. Moreover, our method reduces the cost of computational resource from the exponential growth~\cite{in2014} to the polynomial growth with the particle number. Since the work distribution provides a great deal of information about thermodynamics in a nonequilibrium process, our method will hopefully significantly advance the development of quantum stochastic thermodynamics, especially for quantum many-body systems.

\textit{Two-point measurement scheme}.---In this scheme, the trajectory work for an initial canonical ensemble is determined by the projective measurements over the Hamiltonians before and after the work protocol. Suppose the system evolves from time $t=t_0$ to time $t=t_1$ with the initial state $\hat \rho$ and the time-dependent Hamiltonian $\hat H(t)$. For simplicity, let $\hat \rho$ commute with $\hat H(t_0)$.  We measure the energy of the system at the initial and final time over the Hamiltonians $\hat H_0=\hat H(t_0)$ and $\hat H_1=\hat H(t_1)$ respectively. Also, we assume the states of the system are projected into $|E^0_m\rangle$ and $|E^1_n\rangle$ after the corresponding measurements. Here, $\{|E^0_m\rangle\}$ and $\{|E^1_n\rangle\}$ are the instantaneous orthonormal bases, i.e., $\hat H_0|E^0_m\rangle=E^0_m|E^0_m\rangle$ and $\hat H_1|E^1_n\rangle=E^1_n|E^1_n\rangle$. Then the trajectory work for the above realization is given by $w_{nm}=E^1_n-E^0_m$. After repeating the above procedure with the same canonical ensemble and the same driving protocol numerous times, the work distribution is obtained as follows
\begin{eqnarray}
 P(w)=\sum_{m,n}\langle E^0_m|\hat \rho|E^0_m\rangle |\langle E^1_n|\hat U(t)|E^0_m\rangle|^2\delta(w-w_{nm}),
\end{eqnarray}
where $\hat U(t_1,t_0)=\mathcal T \exp[-i\int_{t_0}^{t_1}\mathrm ds\hat H(s)]$ is the unitary evolution operator (we set $\hbar=1$ in this letter) and $\mathcal T$ is the time-ordered operator. For theoretical analysis, it is a better choice to consider the Fourier transform of the work distribution which is known as the characteristic function of work (CFW) $G(v)$~\cite{wo2007}
%(the Fourier transform of $P(w)$)
\be
\label{e2}
G(v)=\int_{t_0}^{t_1}\mathrm dw P(w)e^{ivw}=\mathrm{Tr}(e^{iv\hat H^H_1}e^{-iv\hat H_0}\hat\rho),
\ee
where the index $H$ in $\hat H^H_1$ indicates that the operator is in Heisenberg picture, i.e., $\hat O^H(t)=\hat U^\dag(t,t_0)\hat O(t)\hat U(t,t_0)$. It is worth mentioning that $G(v)$ encodes the same information as $P(w)$.
%The two-point measurement scheme can also be generalized to a grand ensemble initial state straightforwardly~\cite{}.
%In some  special cases, for example the time-dependent Hamiltonian is in the quadratic polynomial of bosonic operators and the quadratic form of fermionic operators and the system is at the equilibrium initial state, we find it is possible to obtain the analytical expression of $G(v)$ by using the matrix (Lie group) representation technique. We show the details in Sec.~\ref{}.

\textit{Calculating method of the CFW}.---As long as the Hamiltonian is a quadratic polynomial of bosonic or fermionic operators and the initial state is a thermal equilibrium state, the CFW is solvable. In the following, we elucidate our method for arbitrary work protocols.

For later convenience, we introduce some notations first. In a Hilbert space,
%let us define the commutator and the anticommutator of two operators $\bm\cdot$ and $\bm\star$ as $[\bm\cdot, \bm\star]_{\pm}=\bm\cdot\bm\star\pm\bm\star\bm\cdot$. And
let $\hat a_i (\hat b_i)$ and $\hat a_i^\dag (\hat b_i^\dag), i=1,2,\ldots,2n.$ denote the annihilation and creation operators for bosons (fermions). These operators can be considered as the elements of two column vectors, $\bm \alpha=(\hat a_1,\hat a_2,\ldots,\hat a_n,\hat a_1^\dag,\hat a_2^\dag,\ldots,\hat a_n^\dag)^T$ and $\bm \beta=(\hat b_1,\hat b_2,\ldots,\hat b_n,\hat b_1^\dag,\hat b_2^\dag,\ldots,\hat b_n^\dag)^T$, where $T$ denotes the matrix transpose.

Now, without losing generality, we first consider a time-dependent bosonic system  whose Hamiltonian can be written as %(we almost do not show the time dependence explicitly in the following for the concision of symbols)
\be
\label{e3}
\hat{H}(t)=\frac{1}{2}\bm{\alpha}^T A(t) \bm{\alpha}+B^T(t) \bm{\alpha}+C(t),
\ee
where $A(t)$ is a $2n\times2n$ matrix, $B(t)$ is a $2n$-dimensional column vector and $C(t)$ is a scalar. Here, $A(t)$, $B(t)$ and $C(t)$ are work parameters that are controlled by an external agent.
From Eq.~(\ref{e2}), we see that the calculation of the CFW can be decomposed into two steps. The first step is the calculation of $\hat H_1^H$, and the second step is to calculate the trace of the product of several exponential
operators. Based on these facts, the CFW for systems described by the Hamiltonian (Eq.~(\ref{e3})) can be obtained according to the following procedure:

\begin{enumerate}[fullwidth,itemindent=0em,label=(\arabic*)]
\item Solve the time-dependence of $\bm\alpha^H(t)$. Let $\bm\alpha^H(t)=(\hat a_1^H(t),\hat a_2^H(t),\ldots,\hat a_n^H(t),\hat a_1^{\dag H}(t),\hat a_2^{\dag H}(t),\ldots,\hat a_n^{\dag H}(t))^T$ denote $\bm\alpha$ in Heisenberg picture. Then $\bm\alpha^H(t)$ satisfies the following equation% Solve the Heisenberg equation of motion of $\bm\alpha^H(t)=(\hat a_1^H(t),\hat a_2^H(t),\ldots,\hat a_n^H(t),\hat a_1^{\dag H}(t),\hat a_2^{\dag H}(t),\ldots,\hat a_n^{\dag H}(t))^T$,
\be
\label{e17}
\frac{\mathrm d}{\mathrm dt}\hat\alpha_j^H(t)=-i[\hat\alpha_j^H(t),\hat H^H(t)]\ \ \ j=1,2,\cdots,2n.
\ee
Because the Hamiltonian is quadratic, the differential equation (Eq.~(\ref{e17})) is linear. As a result, the solution of $\bm\alpha^H(t)$ can be written in the following linear form% and the solution is also the linear combination of the initial value. Hence, we have
\be
\label{e18}
\bm\alpha^H(t)=D(t)\bm\alpha+E(t),
\ee
where $D(t)$ is the general solution and $E(t)$ is the particular solution of the linear ordinary differential equation (Eq.~({\ref{e17}}))~\cite{od2012}. Substituting Eq.~(\ref{e18}) into Eq.~(\ref{e17}), we obtain
 \begin{gather}
  \begin{split}
\dot{D}(t)&=-i\tau_BA(t)D(t),
 D(t_0)=I,\\
\dot{E}(t)&=-i\tau_B[A(t)E(t)+B(t)],
E(t_0)=0,
  \end{split}
 \end{gather}
where the overhead dot denotes the time derivative, $\tau_B$ is a constant matrix given by
\be
\tau_B=\left(
 \begin{matrix}
 0&\textbf{I}\\
 -\textbf{I}&0
 \end{matrix}
 \right), \nonumber
\ee
%$(\tau_B)_{ij}=[\alpha_i,\alpha_j]_-$,
and $I$ ($\textbf{I}$) denotes the $2n\times 2n$ ($n\times n$) identity matrix. Then by substituting $\bm{\alpha}^H(t_1)=D(t_1)\bm\alpha+E(t_1)$ into $\hat{H}^H_1=\frac{1}{2}\bm{\alpha}^{HT}(t_1) A(t_1) \bm{\alpha}^H(t_1)+B^T(t_1) \bm{\alpha}^H(t_1)+C(t_1)$, we obtain the expression of the Hamiltonian in Heisenberg picture. Please note that the Hamiltonian $\hat H^H_1$ is still a quadratic polynomial of $\bm\alpha$.

\item Calculate the trace of the product of several exponential operators in Eq.~(\ref{e2}) by utilizing the Lie-group representation technique which will be elucidated in next section.
\end{enumerate}

Similarly, we can also consider a time-dependent fermionic system whose Hamiltonian can be written as
\be
\label{e9}
\hat{H}(t)=\frac{1}{2}\bm{\beta}^T A(t) \bm{\beta}+B^T(t) \bm{\beta}+C(t).
\ee
We would like to emphasize that when there is no linear part in Eq.~(\ref{e9}), i.e., $B(t)\equiv0$, the calculating method for quadratic fermionic systems is almost the same as that for quadratic bosonic systems mentioned above. However, when $B(t)\not\equiv 0$, the above method fails, because the Bogoliubov transformation does not apply in this situation~\cite{in2010,no2005,qu1983,di1979}. For this situation, we can employ a new method which is described as follows to bypass this difficulty.

We introduce an artificial Hamiltonian $\hat{\widetilde{H}}(t)$ by adding a pair of  identical fermionic `ghost' operators $\hat b_0$ and $\hat b^\dag_0$~\cite{di1979}
 \begin{gather}
  \begin{split}
\hat{\widetilde{H}}(t)\equiv&\frac{1}{2}\bm{\beta}^T A(t) \bm{\beta}+(\hat b^\dag_0-\hat b_0)B^T(t) \bm{\beta}+C(t)\\
                   =&\frac{1}{2}\widetilde{\bm{\beta}}^T \widetilde{A}(t) \widetilde{\bm{\beta}}+C(t),
  \end{split}
 \end{gather}
where $\widetilde{\bm{\beta}}=(\hat b_0,\hat b_1,\ldots,\hat b_n,\hat b_0^\dag,\hat b_1^\dag,\ldots,\hat b_n^\dag)^T$. The Heisenberg equation of motion for $\widetilde{\bm\beta}^H(t)=(\hat b_0^H(t),\hat b_1^H(t),\ldots,\hat b_n^H(t),\hat b_0^{\dag H}(t),\hat b_1^{\dag H}(t),\ldots,\hat b_n^{\dag H}(t))^T$ is linear again. Now, the artificial Hamiltonian has no linear part. The above method can be applied, and we obtain the CFW $\widetilde{G}(v)$ corresponding to $\hat{\widetilde{H}}(t)$ by similarly following the procedure for bosons. Finally, the CFW corresponding to $\hat{H}(t)$ is obtained by $G(v)=\frac{1}{2}\widetilde{G}(v)$ (see supplemental material).

\textit{Lie-group representation technique}.---A key point in the calculation of the trace (Eq.~(\ref{e2})) is to calculate the product of several exponential operators and this can be significantly simplified by utilizing the Lie-group representation technique. A number of different representations of Lie algebras can be constructed from the bosonic (fermionic) annihilation and creation operators~\cite{no1969,co1983,co1990}. For bosons, we have ($\hat 1$ denotes the identity operator)
\begin{enumerate}[fullwidth,itemindent=0em,label=(\roman*)]
\item The $2n$-dimensional Heisenberg algebra $\mathfrak{h}(n)$
\be
\hat a_i^\dag,\hat a_j,\hat 1\ \ \ i,j=1,2,\ldots,n. \nonumber
\ee
\item The $2n$-dimensional symplectic algebra $\mathfrak{sp}(2n)$
\be
\hat a_i^\dag\hat a_j,\hat a_i^\dag\hat a_j^\dag,\hat a_i\hat a_j\ \ \ i,j=1,2,\ldots,n. \nonumber
\ee
\item The semidirect product of symplectic algebra and Heisenberg algebra $\mathfrak{sp}(2n)\bigotimes_s \mathfrak{h}(n) $
\be
\hat a_i^\dag\hat a_j,\hat a_i^\dag\hat a_j^\dag,\hat a_i\hat a_j,\hat a_i^\dag,\hat a_j,\hat 1\ \ \ i,j=1,2,\ldots,n. \nonumber
\ee
\end{enumerate}
And for fermions, we have
\begin{enumerate}[fullwidth,itemindent=0em,label=(\roman*)]
\setcounter{enumi}{3}
\item The $2n$-dimensional special orthogonal algebra $\mathfrak{so}(2n)$
\be
\hat b_i^\dag\hat b_j-\frac{1}{2}\delta_{ij},\hat b_i^\dag\hat b_j^\dag,\hat b_i\hat b_j\ \ \ i,j=1,2,\ldots,n, \nonumber
\ee
where $\delta_{ij}$ denotes the Kronecker delta function.
\item The $2n+1$-dimensional special orthogonal algebra $\mathfrak{so}(2n+1)$
\be
\hat b_i^\dag\hat b_j-\frac{1}{2}\delta_{ij},\hat b_i^\dag\hat b_j^\dag,\hat b_i\hat b_j,\hat b_i^\dag,\hat b_j,\hat 1\ \ \ i,j=1,2,\ldots,n. \nonumber
\ee
 \end{enumerate}
 The above algebras are all over the complex number field $\mathbb C$. We would like to emphasize that all quadratic Hamiltonians can be constructed by the above Lie algebras~\cite{co1983}. Hence, our study has exhausted \cite{foot1} all situations of quantum systems described by quadratic Hamiltonians.
 The exponential map maps the Lie algebras to the corresponding Lie groups, the Heisenberg group $\mathrm{H}(n)$, the symplectic group $\mathrm{Sp}(2n)$, the semidirect product of the symplectic group and Heisenberg group $\mathrm{Sp}(2n)\bigotimes_s \mathrm{H}(n)$ and the spin groups $\mathrm{Spin}(2n)$ and $\mathrm{Spin}(2n+1)$~\cite{co1983}. Thus, the product (group multiplication) of several exponential quadratic operators with or without a linear part is equal to another exponential quadratic operator with or without a linear part~\cite{no1969}. The trace of the latter can be obtained~\cite{en1978,co2012,an1997} by using the coherent-state representation technique for bosons and fermions. In the following, we give the trace formula corresponding to the product of  several exponential quadratic operators without a linear part.
 \begin{enumerate}[fullwidth,itemindent=0em,label=(\arabic*)]
\item Quadratic bosonic operators without a linear part: $\mathrm{Sp}(2n)$
 \be
 \label{e11}
 \mathrm{Tr}\left[\prod_{i=1}^{m}\mathrm{exp}\left(\frac{1}{2}\bm\alpha^T S_i\bm\alpha\right)\right]=\left\{(-1)^n\mathrm{det}\left[\prod_{i=1}^{m}\mathrm{exp}(\tau_BS_i)-I\right]\right\}^{-\frac{1}{2}},
 \ee
 where $m$ denotes the number of the exponential operators.
\item Quadratic fermionic operators without a linear part: $\mathrm{Spin}(2n)$
 \be
 \label{e12}
 \mathrm{Tr}\left[\prod_{i=1}^{m}\mathrm{exp}\left(\frac{1}{2}\bm\beta^T R_i\bm\beta\right)\right]=\left\{\mathrm{det}\left[\prod_{i=1}^{m}\mathrm{exp}(\tau_FR_i)+I\right]\right\}^{\frac{1}{2}},
 \ee
\end{enumerate}
where
  \be
\tau_F=\left(
 \begin{matrix}
 0&\textbf{I}\\
 \textbf{I}&0
 \end{matrix}
 \right), \nonumber
\ee
$S_i$ ($R_i$) are ($2n\times 2n$) symmetric (antisymmetric) matrices and $\mathrm{exp}(\tau_BS_i)$ ($\mathrm{exp}(\tau_FR_i)$) are the $2n$-dimensional representation matrices of $\mathrm{Sp}(2n)$ ($\mathrm{Spin}(2n)$), acting on the vector space spanned by the basis $\{\alpha_1,\alpha_2,\cdots,\alpha_{2n}\}$ ($\{\beta_1,\beta_2,\cdots,\beta_{2n}\}$). The symmetric (antisymmetric) condition of $S_i$ ($R_i$) is due to the commutation (anticommutation) relation of bosons (fermions) \cite{no1969}. Also, we require the operators in the trace are trace class operators and $\tau_BS_i$ ($\tau_FR_i$) are invertible and diagonalizable~\cite{qu1986}.

For quadratic bosonic operators with a linear part, we can also use the coherent-state representation and Weyl ordering rule~\cite{or1977} to obtain the trace formula:

 \begin{enumerate}[fullwidth,itemindent=0em,label=(\arabic*)]
\setcounter{enumi}{2}
\item Quadratic bosonic operators with a linear part: $\mathrm{Sp}(2n)\bigotimes_s \mathrm{H}(n)$
 \begin{gather}
  \begin{split}
  \label{e13}
 &\mathrm{Tr}\left[\mathrm{exp}\left(\frac{1}{2}\bm\alpha^T S_1\bm\alpha+l_1^T\bm\alpha\right)\mathrm{exp}\left(\frac{1}{2}\bm\alpha^T S_2\bm\alpha+l_2^T\bm\alpha\right)\right]\\
 &=\left\{(-1)^n\mathrm{det}\left[\mathrm{exp}(\tau_BS_1)\mathrm{exp}(\tau_BS_2)-I\right]\right\}^{-\frac{1}{2}}\mathrm{exp}(L),
  \end{split}
 \end{gather}
where $l_1$ and $l_2$ are $2n$-dimensional column vectors. Here $L$ reads
 \begin{gather}
  \begin{split}
L=\sum_{i=1}^2\frac{1}{2}l_i^T[q(\tau_BS_i)-I]S_i^{-1}l_i-\frac{1}{2}l^TS^{-1}l,
  \end{split}
 \end{gather}
where
 \begin{gather}
  \begin{split}
&q(x)=\frac{\mathrm{Tanh}(x)}{x},\\
&S=S_1q(\tau_BS_1)+S_2q(\tau_BS_2),\\
&l^T=l_1^Tq(\tau_BS_1)+l_2^Tq(\tau_BS_2).
  \end{split}
 \end{gather}
 \end{enumerate}

For quadratic fermionic operators with a linear part, it is a bit tricky. By adding a pair of ghost operators (see supplemental material)~\cite{di1979} and using Eq.~(\ref{e12}), we obtain the trace formula

 \begin{enumerate}[fullwidth,itemindent=0em,label=(\arabic*)]
\setcounter{enumi}{3}
\item Quadratic fermionic operators with a linear part: $\mathrm{Spin}(2n+1)\subset\mathrm{Spin}(2n+2)$
 \begin{gather}
  \begin{split}
  \label{e14}
 &\mathrm{Tr}\left[\prod_{i=1}^{m}\mathrm{exp}\left(\frac{1}{2}\bm\beta^T R_i\bm\beta+l_i^T\bm\beta\right)\right]\\
 %&=\frac{1}{2}\mathrm{Tr}[\prod_{i=1}^{m}\mathrm{exp}(\frac{1}{2}\bm\beta^T R_i\bm\beta+(\hat b^\dag_0-\hat b_0)l_i^T\bm\beta)]\\
 &=\frac{1}{2}\left\{\mathrm{det}\left[\prod_{i=1}^{m}\mathrm{exp}(\tau_B\widetilde{R}_i)+I\right]\right\}^{\frac{1}{2}},
  \end{split}
 \end{gather}
where $\widetilde{R}_i$ is determined by the following equation
\be
\frac{1}{2}\widetilde{\bm\beta}^T \widetilde{R}_i\widetilde{\bm\beta}=\frac{1}{2}\bm\beta^T R_i\bm\beta+(\hat b^\dag_0-\hat b_0)l_i^T\bm\beta.
\ee
 \end{enumerate}
We would like to emphasize that the sign of the square root of a complex number in Eqs.~(\ref{e11}--\ref{e13},\ref{e14}) is determined by two properties of the CFW: $G(0)=1$ and $G(v)$ is a smooth function of $v$.
Moreover, some trace formulas corresponding to the subgroups of the Lie groups mentioned above are specials cases of our results. For example, the Levitov's formula~\cite{el1996,an2003} which corresponds to unitary groups~\cite{foot1} can be reproduced from Eqs.~(\ref{e11},\ref{e12}).

\textit{Example}.---In order to illustrate the effectiveness of our method, in the following, we solve the CFW of a 1D generic time-dependent harmonic oscillator which belongs to the group $\mathrm{Sp}(2n)\bigotimes_s \mathrm{H}(n)$ .

Let us consider a generic non-homogeneous second-order linear ordinary differential equation
\be
\label{e20}
\ddot{x}(t)+a(t)\dot{x}(t)+\omega(t)^2x(t)+c(t)=0.
\ee
This equation describes a forced harmonic oscillator with a time-dependent frequency $\omega(t)$ and mass $m(t)$. In addition, it is subject to an external driving force $m(t)c(t)$. From the equation of motion (Eq.~(\ref{e20})), one can derive the Hamiltonian
\be
H(t)=\frac{p^2}{2m(t)}+\frac{1}{2}m(t)\omega^2(t)x^2+m(t)c(t)x.
\ee
Then the parameter $a(t)$ in Eq.~(\ref{e20}) can be expressed as
\be
a(t)=\frac{\mathrm{d}}{\mathrm{d}t}\mathrm{ln}\frac{m(t)}{m(t_0)}. \nonumber
\ee
For simplicity, we set $c(t_0)=0$.

To quantize the system, we employ the canonical commutation relation $[\hat{x},\hat{p}]=i$. Then the solutions of $\hat x^H(t)$ and $\hat p^H(t)$ read
 \begin{gather}
  \begin{split}
\hat{x}^H(t)&=y_1(t)\hat{x}+y_2(t)\frac{\hat{p}}{m(t_0)\omega(t_0)}+f(t),\\
\hat{p}^H(t)&=m(t)\dot{\hat{x}}^H(t),
  \end{split}
 \end{gather}
with
$y_1(t_0)=1,\dot{y_1}(t_0)=0,y_2(t_0)=0,\dot{y_2}(t_0)=\omega(t_0),f(t_0)=0$,
where $y_1(t)$ and $y_2(t)$ are the general solutions and $f(t)=\int_{t_0}^{t}\mathrm dt' [y_1(t)y_2(t')-y_1(t')y_2(t)]\frac{m(t')c(t')}{m(t_0)\omega(t_0)}$ is the particular solution of Eq.~(\ref{e20}). In the derivation, we have used the Abel's identity~\cite{od2012}
\be
\frac{y_1(t)\dot{y}_2(t)-y_2(t)\dot{y}_1(t)}{y_1(t_0)\dot{y}_2(t_0)-y_2(t_0)\dot{y}_1(t_0)}=e^{-\int_{t_0}^{t}\mathrm ds a(s)}=\frac{m(t_0)}{m(t)}.
\ee

Now, let us introduce the ladder operators $\hat{a}=\sqrt{\frac{m_0\omega_0}{2}}(\hat{x}+\frac{i}{m_0\omega_0}\hat{p})$,
$\hat{a}^{\dag}=\sqrt{\frac{m_0\omega_0}{2}}(\hat{x}-\frac{i}{m_0\omega_0}\hat{p})$, where $m_0=m(t_0)$ and $\omega_0=\omega(t_0)$.
Also we define the Fock states as
$\hat a^\dag\hat a\left|n\right\rangle=n\left|n\right\rangle,n=0,1,\ldots$.
Thus, we can rewrite the Hamiltonian into a quadratic polynomial of operators, $\hat a$ and $\hat a^\dag$, and obtain the CFW by the calculating method mentioned above. As for the initial state, we would like to calculate the CFW when the system is initially prepared in a microcanonical state or a canonical equilibrium state. For this purpose, we introduce an artificial density operator $\hat \rho(s)=e^{\mathrm{ln}s\hat a^\dag\hat a}=\sum_ns^n \left|n\right\rangle\left\langle n\right|$ where $s$ is a parameter. Then, the $n$th Fock state (the microcanonical state) $\hat \rho_n=\left|n\right\rangle\left\langle n\right|$ and the canonical equilibrium state $\hat \rho_{th}=e^{-\beta\hat H_0}/Z_{0}$, $Z_{0}=\mathrm{Tr}(e^{-\beta\hat H_0})$ can be generated from $\hat \rho(s)$ by
 \begin{gather}
  \begin{split}
\hat \rho_n&=\frac{1}{n!}\left.\frac{\partial^n}{\partial s^n}\hat \rho_g(s)\right|_{s=0},\\
\hat \rho_{th}&=(1-e^{-\beta\omega_0})\hat \rho(s)|_{s=e^{-\beta\omega_0}},
  \end{split}
 \end{gather}
where $\beta$ is the inverse temperature of the initial thermal state.
\begin{table*}[ht]
 \centering
 \caption{Comparison of results obtained from our method with previous results}
 \label{t1}
 \renewcommand{\arraystretch}{2}
 \begin{tabular}{|c|c|c|}
  \hline
  & Group-theoretical approach & Previous results  \\
  \hline
  \multirow{2}*{}& &Harmonic oscillator with a time-dependent frequency\\
  &Generic time-dependent  &(when $m(t)=m_0, c(t)=0$)~\cite{no2008} \\
  \cline{3-3}
  Boson&harmonic oscillator: $\mathrm{Sp}(2n)\bigotimes_s \mathrm{H}(n)$   &Forced harmonic oscillator \\ &(arbitrary protocol: $m(t)$, $\omega(t)$, $c(t)$)(Eq.~(\ref{e32}))  &(when $m(t)=m_0$, $\omega(t)=\omega_0$)~\cite{sta2008} \\
  \cline{3-3}
  & &Quenched Luttinger Liquid at zero temperature~\cite{ge2012}\\
  \cline{3-3}
  & &A driving quantum scalar field~\cite{wor2019}\\
  \hline
  \multirow{2}*{} &Transverse XY model: $\mathrm{Spin}(2n)$ &Transverse XY model \\
  Fermion&(arbitrary protocol) (Eq.~(S28)) &(sudden quench protocol)~\cite{em2012,st2008,ex2015} \\
  \cline{2-3}
  &Anisotropic XY model in a magnetic field & \\
  &parallel to X- or Y-axis: $\mathrm{Spin}(2n+1)$~\cite{di1979,tw1961} & \\
  \hline
 \end{tabular}
 \label{tab:myfirsttable}
\end{table*}

Using Eq.~(\ref{e13}), we obtain the generating function of the CFW $G(v,s)$ when the initial state is chosen to be $\hat\rho(s)$
\begin{widetext}
 \begin{gather}
  \begin{split}
  \label{e31}
G(v,s)=&\frac{\mathlarger{\exp{ \left\{-\frac{iv m_1c_1^2}{2\omega_1^2}+\frac{iS_{+}\mathrm{cot}(v\omega_1/2)+i(P-S_+R_++S_{-}R_{-})\cot[(i\mathrm{ln}s+v\omega_0)/2]}{\cot^2(v\omega_1/2)-2R_+\cot(v\omega_1/2)\cot[(i\ln s+v\omega_0)/2]+\cot^2[(i\ln s+v\omega_0)/2]}\right\}}}}{\sqrt{2s[-1+\cos(v\omega_1)\cos(i\ln s+v\omega_0)+R_+\sin(v\omega_1)\sin(i\ln s+v\omega_0)]}},
  \end{split}
 \end{gather}
\end{widetext}
with
 \begin{gather}
  \begin{split}
  P=&\frac{2m_1^2\dot f(t_1)}{m_0\omega_0}\left[\frac{c_1}{\omega_1^2}+f(t_1)\right]\left[y_1(t_1)\dot y_1(t_1)+y_2(t_1)\dot y_2(t_1)\right],\\
  R_{\pm}=&\frac{m_1\omega_1}{2m_0\omega_0}\left\{\left[y_1^2(t_1)+y_2^2(t_1)\right]\pm\frac{1}{\omega_1^{2}}\left[\dot{y}_1^2(t_1)+\dot{y}_2^2(t_1)\right]\right\},\nonumber
  \end{split}
 \end{gather}
\be
  S_{\pm}=m_1\omega_1\left\{\left[\frac{c_1}{\omega_1^2}+f(t_1)\right]^2\pm\left[\frac{\dot f(t_1)}{\omega_1}\right]^2\right\},
\ee 
where $m_1=m(t_1)$, $\omega_1=\omega(t_1)$ and $c_1=c(t_1)$.
Thus, the CFW for the microcanonical initial state $G_n(v)$ and for the canonical initial state $G_{th}(v)$ can be generated from $G(v,s)$ by
 \begin{gather}
  \begin{split}
\label{e32}
G_n(v)&=\frac{1}{n!}\left.\frac{\partial^n}{\partial s^n}G(v,s)\right|_{s=0},\\
G_{th}(v)&=(1-e^{-\beta\omega_0})G(v,s)|_{s=\mathrm{e}^{-\beta\omega_0}}.
  \end{split}
 \end{gather}
It is  straightforward to check that the above expressions of the CFW, $G_n(v)$ and $G_{th}(v)$, satisfy the following conditions: (1) $G_n(0)=G_{th}(0)=1$; (2) $G_{th}(i\beta)=Z_1/Z_0$, where $Z_1=\mathrm{Tr}(e^{-\beta\hat H_1})$. The two conditions indicate the normalization condition of the CFW and the validity of Jarzynski equality.
We would like to emphasize that our results (Eq.~(\ref{e32})) include the results of the forced harmonic oscillator~\cite{sta2008} and the result of the harmonic oscillator with a time-dependent frequency~\cite{no2008} as special cases. That is, for a forced harmonic oscillator, $m(t)=m_0$, $\omega(t)=\omega_0$,
%\be
%G_n(v)=\exp[-\frac{iv m_0c_1^2}{2w_0^2}+S_+(e^{iv\omega_0}-1)]L_n[4S_+\sin^2(\frac{v\omega_0}{2})],
%\ee
%and
%\be
%G_{th}(v)=\exp[-\frac{iv m_0c_1^2}{2w_0^2}+S_+(e^{iv\omega_0}-1-\frac{4\sin^2(\frac{v\omega_0}{2})}{e^{\beta\omega_0}-1})],
%\ee
%where $L_n(x)=\sum_{k=0}^n\binom{n}{k}(-x)^k/k!$ denotes the Laguerre polynomial of order $n$, which are consistent with
we reproduce Eqs.~(30,33) in Ref.~\cite{sta2008} ($S_+$ is denoted as $|z|^2$ in that reference); Ror a harmonic oscillator with a time-dependent frequency, $m(t)=m_0, c(t)=0$, we reproduce
%\begin{widetext}
%\be
%G_{th}(v)=\frac{\sqrt{2}\sinh(\frac{\beta\omega_0}{2})}{\sqrt{-1+\cos(v\omega_1)\cos[(v-i\beta)\omega_0]+R_+\sin(v\omega_1)\sin[(v-i\beta)\omega_0}]},
%\ee
%\end{widetext}
%which is exactly
Eq.~(17) in Ref.~\cite{no2008} ($R_+$ is denoted as $Q$ in that reference). As for other examples, the CFW associated with the dynamical Casimir effect is obtained in Ref.~\cite{qu2019} which belongs to the group $\mathrm{Sp}(2n)$. And the CFW of a 1D Transverse XY model is given in the supplemental material which belongs to the group $\mathrm{Spin}(2n)$. In order to demonstrate the powerfulness of our method, we compare the results obtained from the Lie-group representation technique with previous results in TABLE~\ref{t1}.

\textit{Conclusion}.---By extending the matrix representation technique introduced in Refs.~\cite{no1969,en1978}, we invent a universal method based on Lie-group theory to calculate the trace of products of several exponential quadratic operators. Our method is valid as long as the Hamiltonian can be written in the  form of  the quadratic polynomial of bosonic or fermionic operators which is ubiquitous in physics especially in quantum many-body systems. Our trace formulas exhausted all possible Lie groups for quadratic Hamiltonians. Hence, it is the maximum extension of Levitov's formula~\cite{el1996,an2003}. By utilizing these trace formulas, we develop a systematic and general procedure of calculating the work distributions in many quantum systems under arbitrary work protocols in a much more efficient way. In comparison with previous results which are restricted to either several specific models or very special protocols, we can obtain the analytical expression of the CFW by a unified method. In addition, our method reduces the cost of the computational resource from the exponential growth (the dimension of Hilbert space) to the polynomial growth (the dimension of the representation space of Lie group) with the particle number. As examples, we calculate the CFW of a generic time-dependent harmonic oscillator and the CFW of a time-dependent transverse XY model (see supplemental material) to illustrate the powerfulness of our method. We also would like to emphasize that this method has potential applications in many other fields. Such as quantum optics and statistical physics. Examples include the partition functions, Out-of-time-ordered correlation functions~\cite{ch2018,th2017,ou2019}, Loschmidt echos~\cite{en2001,de2006}, Full-Counting Statistics~\cite{fu2015,qua2019} and other correlation functions. Studies about applications of our method in these problems will be given in forthcoming papers.
\begin{acknowledgments}
H. T. Quan gratefully acknowledges support from
the National Science Foundation of China under grants
11775001, 11534002, and 11825001.
\end{acknowledgments}

  \end{document}

%\preprint{}

\title{Supplemental Material: Group-theoretical approach to the calculation of quantum work distribution}


\author{Zhaoyu Fei}
%\email[]{kenfuno@pku.edu.cn}
%\thanks{}
\affiliation{School of Physics, Peking University, Beijing 100871, China}
\author{H. T. Quan} \email[Email: ]{htquan@pku.edu.cn}
\affiliation{School of Physics, Peking University, Beijing 100871, China}
\affiliation{Collaborative Innovation Center of Quantum Matter, Beijing 100871, China}

\maketitle

\section*{APPENDIX A: The ghost operator method for fermionic system}

Let $\mathcal{H}$ denote the $2^n$-dimensional Hilbert space of a fermionic system with the basis vectors $|\{n_i\}\rangle\equiv\otimes_i|n_i\rangle,n_i=0,1$ and the ladder operators $\hat b_j,\hat b_j^\dag$ ($i,j=1,2,\ldots,n$). We introduce a $2^{n+1}$-dimensional Hilbert space $\widetilde{\mathcal H}$ by adding a pair of fermionic ghost operators $\hat b_0,\hat b_0^\dag$ with the basis vectors $|0\rangle,|1\rangle$. These operators obey the anticommutation relation
 \begin{gather}
  \begin{split}
  \label{e29}
&\{\hat b_i,\hat b^\dag_j\}=\delta_{ij}\\
&\{\hat b_i,\hat b_j\}=\{\hat b^\dag_i,\hat b^\dag_j\}=0\ \ \ \ i,j=0,1,\ldots,n,
  \end{split}
 \end{gather}
where $\{\hat U,\hat V\}=\hat U\hat V+\hat V\hat U$.
Then, the Hilbert space $\widetilde{\mathcal H}$ can be divided into two $2^n$-dimensional subspaces $\widetilde{\mathcal H}=\widetilde{\mathcal H}_+\oplus\widetilde{\mathcal H}_-$ with the basis vectors $|\{n_i\}\rangle_+\oplus|\{n_i\}\rangle_-$, where
\be
|\{n_i\}\rangle_\pm=\frac{1}{\sqrt{2}}(|0\rangle\pm|1\rangle)\otimes|\{n_i\}\rangle.
\ee
Correspondingly, the orthogonal projective operators on the two subspaces are $\hat P_\pm=\frac{1}{2}[\hat 1\pm(\hat b_0+\hat b_0^\dag)]$, where $\hat 1$ denotes the identity operator. According to Ref.~\cite{di1979}, we construct an isomorphic mapping between the Hilbert space $\mathcal H$ and the subspace $\widetilde{\mathcal H}_+$ in the following way:
 \begin{gather}
  \begin{split}
 |\{n_i\}\rangle&\leftrightarrow|\{n_i\}\rangle_+\\
 \hat b_j&\leftrightarrow(\hat b_0^\dag-\hat b_0)\hat b_j\\
 \hat b_j^\dag&\leftrightarrow(\hat b_0^\dag-\hat b_0)\hat b_j^\dag\\
 f(\{\hat b_j\},\{\hat b_j^\dag\})&\leftrightarrow f(\{(\hat b_0^\dag-\hat b_0)\hat b_j\},\{(\hat b_0^\dag-\hat b_0)\hat b_j^\dag\})\ \ \ \ i,j=1,2,\ldots,n,
  \end{split}
 \end{gather}
 where $f(\{\hat b_j\},\{\hat b_j^\dag\})\equiv f(\hat b_1,\hat b_2,\ldots,\hat b_n,\hat b^\dag_1,\hat b^\dag_2,\ldots,\hat b^\dag_n)$ is an arbitrary smooth function of $\hat b_j$ and $\hat b_j^\dag$. Hence, if $f(\{\hat b_j\},\{\hat b_j^\dag\})$ is trace-class, we can calculate $\mathrm{Tr}[f(\{\hat b_j\},\{\hat b_j^\dag\})]$ both in $\mathcal H$ and in $\widetilde{\mathcal H}$:
 \begin{gather}
  \begin{split}
 \mathrm{Tr}[f(\{\hat b_j\},\{\hat b_j^\dag\})]&=\sum_{\{n_i\}}\langle\{n_i\}|f(\{\hat b_j\},\{\hat b_j^\dag\})|\{n_i\}\rangle\\
 &=\sum_{\{n_i\}}{}_+\langle\{n_i\}|f(\{(\hat b_0^\dag-\hat b_0)\hat b_j\},\{(\hat b_0^\dag-\hat b_0)\hat b_j^\dag\})|\{n_i\}\rangle_+\\
 &=\sum_{n_0,\{n_i\}}\langle n_0|\otimes\langle\{n_i\}|\hat P_+f(\{(\hat b_0^\dag-\hat b_0)\hat b_j\},\{(\hat b_0^\dag-\hat b_0)\hat b_j^\dag\})\hat P_+|n_0\rangle\otimes|\{n_i\}\rangle\\
 &=\mathrm{Tr}[\hat P_+f(\{(\hat b_0^\dag-\hat b_0)\hat b_j\},\{(\hat b_0^\dag-\hat b_0)\hat b_j^\dag\})\hat P_+]\\
 &=\mathrm{Tr}[\hat P_+f(\{(\hat b_0^\dag-\hat b_0)\hat b_j\},\{(\hat b_0^\dag-\hat b_0)\hat b_j^\dag\})],
  \end{split}
 \end{gather}
 where we have used the property $\hat P^2_\pm=\hat P_\pm$. Finally, because $f(\{(\hat b_0^\dag-\hat b_0)\hat b_j\},\{(\hat b_0^\dag-\hat b_0)\hat b_j^\dag\})$ commutes with the parity operator $\hat \Pi=\exp{(i\pi\sum_{i=0}^n \hat b_i^\dag\hat b_i)}$, we have
 \begin{gather}
  \begin{split}
 \mathrm{Tr}[\hat P_+f(\{(\hat b_0^\dag-\hat b_0)\hat b_j\},\{(\hat b_0^\dag-\hat b_0)\hat b_j^\dag\})]&=\mathrm{Tr}[\hat \Pi^{-1}\hat\Pi\hat P_+f(\{(\hat b_0^\dag-\hat b_0)\hat b_j\},\{(\hat b_0^\dag-\hat b_0)\hat b_j^\dag\})]\\
 &=\mathrm{Tr}[\hat{\Pi}\hat P_+\hat{\Pi^{-1}}f(\{(\hat b_0^\dag-\hat b_0)\hat b_j\},\{(\hat b_0^\dag-\hat b_0)\hat b_j^\dag\})]\\
 &=\mathrm{Tr}[\hat P_-f(\{(\hat b_0^\dag-\hat b_0)\hat b_j\},\{(\hat b_0^\dag-\hat b_0)\hat b_j^\dag\})]\\
 &=\frac{1}{2}\mathrm{Tr}[(\hat P_++\hat P_-)f(\{(\hat b_0^\dag-\hat b_0)\hat b_j\},\{(\hat b_0^\dag-\hat b_0)\hat b_j^\dag\})]\\
 &=\frac{1}{2}\mathrm{Tr}[f(\{(\hat b_0^\dag-\hat b_0)\hat b_j\},\{(\hat b_0^\dag-\hat b_0)\hat b_j^\dag\})].
  \end{split}
 \end{gather}

\section*{APPENDIX B: 1D Transverse XY model}

Let us consider a 1D Transverse XY model described by the following Hamiltonian~\cite{tw1961,qu2012,st1962,de2006,de2006}
\be
\hat H=-\frac{J}{2}\sum_{l=1}^{N}[(1+\gamma)\hat \sigma_l^x\hat \sigma_{l+1}^x+(1-\gamma)\hat \sigma_l^y\hat \sigma_{l+1}^y]-\Gamma\sum_{l=1}^N\hat \sigma_l^z,
\ee
where $x,y,z$ are the Cartesian coordinates, $\hat\sigma_{l}^{x,y,z}$ are Pauli operators, $J$ is the coupling strength, $\gamma$ is the anisotropic parameter and $\Gamma$ denotes the strength of the external magnetic field in the $z$ direction. Also, we assume the Born-von K\'{a}rm\'{a}n cyclic condition and the lattice points are labeled with $l=1,2,\ldots,N,\ N+l\equiv l$. We assume $N$ to be a even number for simplicity.

As usual, we use the Jordan-Wigner transformation~\cite{tw1961,qu2012,st1962,dy2005} to simplify the system,
\be
\hat \sigma^z_l=2\hat b_l^\dag\hat b_l-1,
\ee
\be
\hat \sigma^x_l=(\hat b^\dag_l+\hat b_l)\prod_{m<l}(1-2\hat b_m^\dag\hat b_m),
\ee
\be
\hat \sigma^y_l=-i(\hat b^\dag_l-\hat b_l)\prod_{m<l}(1-2\hat b_m^\dag\hat b_m).
\ee
For later convenience, we define the parity operator $\hat \Pi=\mathrm{exp}(i\pi\sum_{l=1}^N\hat b_l^\dag\hat b_l)=\prod_{l=1}^N(1-2\hat b_l^\dag\hat b_l)$ and the corresponding projectors $\hat P^{\pm}=\frac{1}{2}(1\pm\hat \Pi)$ on the subspaces in which the even ($+$) or odd ($-$) number of lattices are occupied. We would like to emphasize that the parity operator is a conserved quantity $[\hat \Pi,\hat H]=0$ even when the parameters in the Hamiltonian are time-dependent.
Then, we transform the XY chain into a spinless fermionic system~\cite{tw1961,qu2012,st1962,dy2005}
\be
\hat H=\hat P^+\hat H^++\hat P^-\hat H^-,
\ee
where
 \begin{gather}
  \begin{split}
\hat H^{\pm}=&-\Gamma\sum_{l=1}^N(\hat b_l^\dag\hat b_l-\hat b_l\hat b_l^\dag)\\
&-\gamma J\sum_{l=1}^N(\hat b_l^\dag\hat b_{l+1}^\dag-\hat b_l\hat b_{l+1})-J\sum_{l=1}^N(\hat b_l^\dag\hat b_{l+1}-\hat b_l\hat b_{l+1}^\dag)
  \end{split}
 \end{gather}
are the reduced Hamiltonians. And $\hat b_l$'s in $\hat H^+$ ($\hat H^-$) satisfy anti-periodic boundary conditions $\hat b_{l+N}=-\hat b_l$ (periodic boundary conditions $\hat b_{l+N}=\hat b_l$). Furthermore, the Hamiltonian can be diagonalized by the discrete Fourier transform~\cite{dy2005}
\be
\hat b_l=\frac{e^{-i\pi/4}}{\sqrt{N}}\sum_{k\in K^{\pm}}\hat b_ke^{ikl},
\ee
 \begin{gather}
  \begin{split}
\hat H^{\pm}=&-\sum_{k\in K^{\pm}}(\Gamma+J\mathrm{cos}k)(\hat b_k^\dag\hat b_k-\hat b_k\hat b_k^\dag)\\
&-\gamma J\sum_{k\in K^{\pm}}\mathrm{sin}k(\hat b_k^\dag\hat b_{-k}^\dag+\hat b_{-k}\hat b_{k}),
  \end{split}
 \end{gather}
where $K^+=\{\pi n/N\},n=1-N,3-N,\ldots,N-1$. and $K^-=\{\pi n/N\},n=2-N,4-N,\ldots,N$,
and the Bogoliubov transformation~\cite{em2012}
\be
\hat b_{\pm k}=\hat \zeta_{\pm k}\mathrm{cos}(\frac{\phi_k}{2})\pm\hat \zeta^\dag_{\mp k}\mathrm{sin}(\frac{\phi_k}{2}),
\ee
where
 \begin{gather}
  \begin{split}
\mathrm{cos}(\phi_k)=&-\frac{2(\Gamma+J\mathrm{cos}k)}{\omega_k}\\
\mathrm{sin}(\phi_k)=&\frac{2\gamma J\mathrm{sin}k}{\omega_k},
  \end{split}
 \end{gather}
with $\omega_k=2\sqrt{(\Gamma+J\mathrm{cos}k)^2+(\gamma J\mathrm{sin}k)^2}$ being the eigenenergy when $k\neq0,\pi$. When $k=0,\pi$, $\phi_k=0$ and $w_k=-2(\Gamma+J),-2(\Gamma-J)$.
Finally, we obtain the diagonal Hamiltonian with the quasiparticle operators $\hat \zeta_k$ and $\hat \zeta^\dag_k$
\be
\hat H^{\pm}=\sum_{k\in K^{\pm}}\omega_k(\hat \zeta^\dag_k\hat\zeta_k-\frac{1}{2}).
\ee
If $N$ is odd, the results are the same except $K^+=\{\pi n/N\},n=2-N,4-N,\ldots,N$. and $K^-=\{\pi n/N\},n=1-N,2-N,\ldots,N-1$.

When the parameter $J(t),\gamma(t)$ and $\Gamma(t)$ are time-dependent, the vector $\bm\xi_k=(\hat b_k,\hat b_{-k},\hat b_k^\dag, \hat b_{-k}^\dag)$ is time-independent and the vector $\bm\eta_k(t)=(\hat \zeta_k(t),\hat \zeta_{-k}(t),\hat \zeta_k^\dag(t), \hat \zeta_{-k}^\dag(t))$ is time-dependent. Also, we almost do not show the time dependence explicitly in the following for the concision of symbols. Thus, the Hamiltonian can be rewritten with $\bm\eta_k$
\be
\hat H^{\pm}=\frac{1}{2}\sum_{k\in K^\pm\atop k\geq0}\bm\eta_k^T\Lambda_k\bm\eta_k,
\ee
where
\be
\Lambda_k= \left(\begin{matrix}
 0&0&\omega_k&0\\
 0&0&0&\omega_k\\
 -\omega_k&0&0&0\\
 0&-\omega_k&0&0
 \end{matrix}\right),
\ee
when $k\neq 0,\pi$. When $k=0,\pi$,
\be
\Lambda_k= \left(\begin{matrix}
 0&\omega_k\\
 -\omega_k&0
 \end{matrix}\right).
\ee
Alternatively, the Hamiltonian can be rewritten with $\bm\xi_k$
\be
\hat H^{\pm}=\frac{1}{2}\sum_{k\in K^\pm\atop k\geq0}\bm\xi_k^T\Omega^T_k \Lambda_k \Omega_k\bm\xi,
\ee
where when $k\neq 0,\pi$,
\be
\Omega_k= \left(\begin{matrix}
 \mathrm{cos}(\frac{\phi_k}{2})&0&0&-\mathrm{sin}(\frac{\phi_k}{2})\\
 0&\mathrm{cos}(\frac{\phi_k}{2})&\mathrm{sin}(\frac{\phi_k}{2})&0\\
 0&-\mathrm{sin}(\frac{\phi_k}{2})&\mathrm{cos}(\frac{\phi_k}{2})&0\\
 \mathrm{sin}(\frac{\phi_k}{2})&0&0&\mathrm{cos}(\frac{\phi_k}{2})
 \end{matrix}\right)
\ee
is the rotation matric between the two vectors. When $k=0,\pi$,
\be
\Omega_k= \left(\begin{matrix}
 1&0\\
 0&1
 \end{matrix}\right).
\ee

Next, when $k\neq 0,\pi$, we solve the Heisenberg equations of motion of $\bm\xi_k^H(t)$
 \begin{gather}
  \begin{split}
  \label{e23}
\frac{\mathrm d}{\mathrm dt}\bm{\xi}^H_k(t)&=-i[\bm{\xi}^H_k(t),H^\pm(t)]\\
&=-i\left(
 \begin{matrix}
 0&0&1&0\\
 0&0&0&1\\
 1&0&0&0\\
 0&1&0&0
 \end{matrix}
 \right)\Omega^T_k(t) \Lambda_k(t) \Omega_k(t)\bm{\xi}^H_k(t)\ \ \ \ k\in K^\pm,
  \end{split}
 \end{gather}
by assuming the solution is $\bm\xi_k^H(t)=\Xi_k(t)\bm\xi_k$
\be
\Xi_k(t)=\left(\begin{matrix}
 x_k^{11}(t)&0&0&x_k^{12}(t)\\
 0&x_k^{11}(t)&-x_k^{12}(t)&0\\
 0&-x_k^{21}(t)&x_k^{22}(t)&0\\
 x_k^{21}(t)&0&0&x_k^{22}(t)
 \end{matrix}\right).
\ee
Then the time-dependent matrix $X_k=(x_k^{ij})_{2\times2}$ satisfies
\be
\frac{\partial}{\partial t} X_k=-i\omega_k\left(\begin{matrix}
 \mathrm{cos}(\phi_k)&-\mathrm{sin}(\phi_k)\\
 -\mathrm{sin}(\phi_k)&-\mathrm{cos}(\phi_k)
 \end{matrix}\right)X_k,X_k(t_0)= \left(\begin{matrix}
 1&0\\
 0&1
 \end{matrix}\right)
\ee
due to Eq.~(\ref{e23}).
Also, the Heisenberg equation of motion of
$\bm\eta_k^H(t)=\Upsilon_k(t)\bm\eta_k$ is solved by
\be
\Upsilon_k(t)=\left(\begin{matrix}
 y_k^{11}(t)&0&0&y_k^{12}(t)\\
 0&y_k^{11}(t)&-y_k^{12}(t)&0\\
 0&-y_k^{21}(t)&y_k^{22}(t)&0\\
 y_k^{21}(t)&0&0&y_k^{22}(t)
 \end{matrix}\right),
\ee
where
\be
Y_k(t)=
(y_k^{ij}(t))_{2\times2}=\left(\begin{matrix}
 \mathrm{cos}(\frac{\phi_k(t)}{2})&-\mathrm{sin}(\frac{\phi_k(t)}{2})\\
 \mathrm{sin}(\frac{\phi_k(t)}{2})&\mathrm{cos}(\frac{\phi_k(t)}{2})
 \end{matrix}\right)X_k(t)\left(\begin{matrix}
 \mathrm{cos}(\frac{\phi_k(t_0)}{2})&\mathrm{sin}(\frac{\phi_k(t_0)}{2})\\
 -\mathrm{sin}(\frac{\phi_k(t_0)}{2})&\mathrm{cos}(\frac{\phi_k(t_0)}{2})
 \end{matrix}\right).
\ee
%Here,
%\be
%R_k=\left(\begin{matrix}
% \mathrm{cos}(\frac{\phi_k}{2})&-\mathrm{sin}(\frac{\phi_k}{2})\\
% \mathrm{sin}(\frac{\phi_k}{2})&\mathrm{cos}(\frac{\phi_k}{2})
% \end{matrix}\right)
%\ee
%is the rotation matrix between $X_k$ and $Y_k$.
%Then, the solution of the Heisenberg equation of motion reads
%$\bm\eta_k^H(t)=C_k(t)\bm\eta_k$
%where
%\be
%C_k(t)=\left(\begin{matrix}
% y_k^{11}(t)&0&0&y_k^{12}(t)\\
% 0&y_k^{11}(t)&-y_k^{12}(t)&0\\
% 0&-y_k^{21}(t)&y_k^{22}(t)&0\\
% y_k^{21}(t)&0&0&y_k^{22}(t)
% \end{matrix}\right).
%\ee
%or when $k=0,\pi$
%\be
%\tau_B=\left(
% \begin{matrix}
% 0&1\\
% 1&0
% \end{matrix}
% \right).
%\ee
Moreover, when $k=0,\pi$, we have $X_k(t)=Y_k(t)=\Xi_k(t)=\Upsilon_k(t)$.

Finally, using Eq.~(10) in the main text, we obtain the CFW of the transverse XY chain $G(v)$ with a canonical initial state
 \begin{gather}
  \begin{split}
  \label{es28}
G(v)=&\frac{\mathrm{Tr}[e^{iv\hat H^H_1}e^{(-iv-\beta)\hat H_0}]}{\mathrm{Tr}(e^{-\beta\hat H_0})}\\
    =&\frac{\mathrm{Tr}[\hat{P}_+e^{iv\hat H^{+H}_1}e^{(-iv-\beta)\hat H^+_0}]+\mathrm{Tr}[\hat{P}_-e^{iv\hat H^{-H}_1}e^{(-iv-\beta)\hat H^-_0}]}{\mathrm{Tr}(\hat{P}_+e^{-\beta\hat H^+_0})+\mathrm{Tr}(\hat{P}_-e^{-\beta\hat H^-_0})}\\
    =&g(v)/g(0),
  \end{split}
 \end{gather}
with
\be
g(v)=\prod_{k\in K^{+}}g^{+}_k(v)+\prod_{k\in K^{+}}g^{-}_k(v)+\prod_{k\in K^{-}}g^{+}_k(v)-\prod_{k\in K^{-}}g^{-}_k(v),
\ee
where $\hat H_0^\pm=\hat H^\pm(t_0)$, $\hat H_1^{\pm H}=\hat H^{\pm H}(t_1)$ and
 \begin{gather}
  \begin{split}
  \label{es30}
g^{\pm}_k(v)=&\{\pm1+\mathrm{cos}(v\omega_k^1)\mathrm{cos}[(v-i\beta)\omega_k^0]\\
&+Q_k\mathrm{sin}(v\omega_k^1)\mathrm{sin}[(v-i\beta)\omega_k^0]\}^{\frac{1}{2}},
  \end{split}
 \end{gather}
with $\omega_k^0=\omega_k(t_0)$, $\omega_k^1=\omega_k(t_1)$
and
\begin{eqnarray}
Q_k=2y_k^{11}(t_1)y_k^{22}(t_1)-1.
\end{eqnarray}
Also, the above expression of the CFW $G(v)$ satisfies the normalization condition of the CFW, $G(0)=1$ and Jarzynski equality, $G(i\beta)=Z_1/Z_0$.
If the number of the lattice is very large, i.e., $N\rightarrow \infty$, the energy spectral of the two subspaces will be the same~\cite{st1962,re1976}. Hence we have
\be
\lim_{N\rightarrow\infty}\frac{1}{N}\ln G(v)=\frac{1}{\pi}\int_0^\pi\mathrm{d}k\ln\frac{g_k^+(v)}{g_k^+(0)}.
\ee
Also, we obtain the first two moments of work
\be
\langle w\rangle=-i\left. \frac{\partial \ln G}{\partial v} \right|_{v=0}=\frac{N}{2\pi}\int_0^\pi\mathrm dk(\omega_k^0-Q_k\omega_k^1)\tanh{(\frac{\beta\omega_k^0}{2})},
\ee
 \begin{gather}
  \begin{split}
\sigma_w^2=&-\left.\frac{\partial^2 \ln G}{\partial v^2} \right|_{v=0}=\frac{N}{4\pi}\int_0^\pi\mathrm dk[(\omega_k^0-Q_k\omega_k^1)^2\\
&+(1-Q_k^2)(\omega_k^1)^2\cosh(\beta\omega_k^0)]\mathrm{sech}^2(\frac{\beta\omega_k^0}{2}).
  \end{split}
 \end{gather}
 %which are also consistent with the results in Ref.~\cite{em2012}.

We would like to emphasize that from our main results Eq.~(\ref{es28}), one can derive the CFW in the sudden quench limit and quantum adiabatic limit:
\begin{enumerate}
\item Sudden quench limit
\be
X_k(t_1)= \left(\begin{matrix}
 1&0\\
 0&1
 \end{matrix}\right).
\ee
Hence, we have $Q_k=\cos[\phi_k(t_1)-\phi_k(t_0)]$. Then, the expression $\prod_{k\in K^{+}}g^{+}_k(v)$ is exactly Eq.~(11) in Ref.~\cite{em2012} which is the CFW of the 1D Transverse XY model in the sudden quench limit without considering the negative-parity Hamiltonian.
\item Quantum adiabatic limit ($N$ is finite)
\be
Y_k(t_1)= \left(\begin{matrix}
 e^{-i\int_{t_0}^{t_1}\mathrm dt \omega_k(t)}&0\\
 0&e^{i\int_{t_0}^{t_1}\mathrm dt \omega_k(t)}
 \end{matrix}\right).
\ee
 Hence, we have $Q_k=1$, which is an indication of quantum adiabaticity (see Eq.~(\ref{es30})). %Actually, the adiabatic limit is broken when the transition probability $\exp(-2\pi^3\gamma J\tau_Q/N^2)$ is big enough, where $J(t)=J,\gamma(t)=\gamma,\Gamma(t)=\Gamma-Jt/\tau_Q$~\cite{dy2005}.
\end{enumerate}